\documentstyle[12pt]{article}

\begin{document}

\begin{flushright}
IMSc/2014/09/08
\end{flushright} 

\vspace{2mm}

\vspace{2ex}

\begin{center}

{\large \bf Massive Compact Objects in} \\

\vspace{2ex}

{\large \bf a Quantum Theory of Gravity } \\

\vspace{6ex}

{\large  S. Kalyana Rama}

\vspace{3ex}

Institute of Mathematical Sciences, C. I. T. Campus, 

Tharamani, CHENNAI 600 113, India. 

\vspace{1ex}

email: krama@imsc.res.in \\ 

\end{center}

\vspace{6ex}

\centerline{ABSTRACT}
\begin{quote} 

A massive compact object is that which forms when a sufficiently
massive star collapses. This is commonly taken to be a black
hole with a singularity surrounded by a horizon and which
evolves by emitting Hawking radiation. In a quantum theory of
gravity, singularities are expected to be resolved and the
evolutions are expected to be unitary. Assuming that such a
theory with these properties exists, and with a few more
physically motivated assumptions, we argue that a massive
compact object has no singularity (by assumption) and must also
have no horizon; otherwise, there may be a loss of
predictability in the case of a black hole candidate observed
today. With no singularity and also with no horizon, the massive
compact object will then evolve as a standard quantum system
with large number of interacting degrees of freedom.

\end{quote}

\vspace{2ex}

%PACS numbers: 04.20.Dw, 04.60.-m, 04.70.Dy, 11.25.-w, 

%PACS numbers: 04.20.Dw Singularities and cosmic censorship

%PACS numbers: 04.60.-m Quantum gravity

%PACS numbers: 04.60.Pp Loop quantum gravity, quantum geometry,
% spin foams

%PACS numbers: 04.70.Dy Quantum aspects of black holes,
% evaporation, thermodynamics

%PACS numbers: 11.25.-w Strings and branes

%PACS numbers: 98.80.Cq Particle-theory and field-theory models
%of the early Universe (including cosmic pancakes, cosmic
%strings, chaotic phenomena, inflationary universe, etc.)

%PACS numbers: 

\newpage

\centerline{\bf I. Introduction}

\vspace{4ex}

A sufficiently massive star will collapse and form what is
referred to as a massive compact object (MCO). This is commonly
taken to be a black hole which has a singularity surrounded by a
horizon and which evolves further by emitting Hawking radiation.
The constituents of the collapsed star are taken to have
disappeared into the singularity. 

In a quantum theory of gravity, singularities are expected to be
resolved, and the process of the collapse of a massive star and
its further evolution are expected to be unitary. In such a
theory, it may be that the MCO formed in the collapse of a
massive star has no singularity but has a horizon, and it
evolves further by emitting Hawking radiation. Or, it may be
that the resulting MCO has no singularity and also has no
horizon, and it evolves further as a standard quantum system
with large number of interacting degrees of freedom. See \cite{
morgan} -- \cite{ bk} for a sample of possibile scenario.

To deduce the nature of the MCOs, it is necessary to understand
the physics of singularities and their resolutions. Although
there are many candidates for a quantum theory of gravity, such
as string theory, quantum Einstein gravity, loop quantum
gravity, and spin network theory, none of them is developed well
enough to be applicable for collapsing massive stars. At
present, therefore, quantitative approaches to such situations
are beyond our reach and only qualitative, physically motivated
approaches seem possible \cite{g, ab, h1, visser, k12,
kkmonopole, rv, bk}

We have found a qualitative, physically motivated approach which
enables one to argue that an MCO formed in the collapse of a
massive star has no singularity (by assumption) and must also
have no horizon; otherwise, there will be a loss of
predictability. See \cite{k12} for an earlier report. With no
horizon, the resulting MCO will evolve further as a standard
quantum system with large number of interacting degrees of
freedom.

In this paper, we describe our approach, arguments, and their
implications leading to the above conclusion. First, we assume
that a quantum theory of gravity exists where singularities are
resolved and the evolutions are unitary. Next, we estimate the
size of the region of singularity resolution, referred to as
`singularity cloud'. If this size is smaller than the horizon
size of a black hole then it can be taken to mean that the MCO
formed in the collapse of a massive star has no singularity but
has a horizon; if larger then the resulting MCO has no
singularity and also has no horizon.

For the purposes of illustration, we present several plausibile
scenario of singularity resolution. The estimated size of the
singularity cloud is smaller than the horizon size in some of
them, and larger in others.

Let the size of the singularity cloud be smaller than the
horizon size. We argue that if this is the case then, for an MCO
observed today which can be thought of as a black hole
candidate, we may not be able to predict when the size of the
singularity cloud inside will exceed its horizon size. This is a
loss of predictabilty, and is new. This is absent in general
relativity theory where the size of the singularity cloud is
always taken to be Planckian. Therefore, if the singularity
resolution is expected in a quantum theory of gravity, but not
the loss of predictability, then it logically follows that the
physics of singularity resolution must be such that the size of
singularity cloud is larger than the horizon size.

Note that an MCO having no singularity and no horizon is likely
to follow also from Mathur's fuzz ball proposal \cite{fuzz} in
string theory applied to a collapsing massive star. In this
paper, we argue that such an MCO must follow in any quantum
theory of gravity obeying the present assumptions.

Note also that, over the years, several works have argued that
the size of the region of singularity resolution must be much
larger than Planck length, see \cite{ morgan, g, rv} for
example. However, in all these works, a horizon is always
assumed to exist when an MCO forms. To the best of our
knowledge, a further logical step is taken for the first time
here which enables one to argue that the size of the singularity
cloud is larger than the horizon size, thus an MCO formed has no
horizon also.

The organisation of this paper is as follows. 

In section {\bf II}, we describe a few aspects of singularity
resolution as needed here. For the purposes of illustration, we
then present several plausibile scenario of singularity
resolution and estimate the size of the singularity cloud. We
state our assumptions at the relevant stages of the paper, but
will collect them together in the final section. In section {\bf
III}, we describe the implications of the size of the
singularity cloud. In section {\bf IV}, we describe the loss of
predictabilty that will result if the singularity cloud is
smaller than horizon in size, and arrive at our main result. In
section {\bf V}, we give a brief summary, discuss a few points,
and close by mentioning two studies that may be pursued
fruitfully. In Appendix A, we give a brief explanation regarding
the meaning of the word predictability as used in this paper.

\vspace{4ex}

\centerline{ \bf II. Aspects of singularity resolution}

\vspace{4ex}

When a star is sufficiently massive, it begins to collapse and
ultimately forms what is referred to as a massive compact object
(MCO). It is believed that the resulting MCO is a black hole:
that the collapse results in the formation of a singularity
surrounded by a horizon and the constituents of the star are
taken to have disappeared into the singularity. In a quantum
theory of gravity such as string theory, quantum Einstein
gravity, loop quantum gravity, and spin network theory, such
singularities are expected to be resolved and the entire process
of collapse, formation of an MCO, and its further evolution are
expected to be unitary.

In this paper, we assume that a quantum theory of gravity exists
where singularities are resolved and the evolutions are unitary.
By assumption, therefore, an MCO will have no singularity. But
it may or may not have a horizon. It will evolve further by
emitting Hawking radiation if there is a horizon, or it will
evolve further as a standard quantum system with large number of
interacting degrees of freedom if there is no horizon.

Let the space be $d$ dimensional, and the spacetime $d + 1$
dimensional, with $d$ taken to be $\ge 3 \;$. Let $l_f$ and $m_f
= l_f^{- 1}$ be the fundamental length and mass scales of the
quantum theory of gravity under consideration. In string theory,
$l_f$ is the string length scale. In other theories, $l_f$ is
the Planck length. Also, we assume that the mass $M$ of the
collapsing star is $\gg m_f$ and that any dimensionless
parameters which may be present in the theory, for example
string coupling constant $g_s$ in string theory, take generic
values of ${\cal O}(1)$ independently of $M \;$.

Quantum gravity effects may be expected to play a role when the
constituents of the collapsing massive star reach a density of
${\cal O}( m_f^{d + 1} ) \;$ or even earlier, see below.
However, these effects are not known in detail. Therefore, as
seems physically reasonable, we assume here that the
singularities in the collapse of a massive star are resolved due
to quantum gravity effects by transforming the star's
constituents into fundamental units, \footnote{ The nature of
fundamental units depends on the specific quantum theory of
gravity. For example, these units are likely to be highly
excited interacting strings, the `fuzz' or, equivalently, the
Kaluza -- Klein monopoles of the fuzzball picture, or the $D0$
branes of the matrix description in string theory; or, loops in
the loop quantum gravity; or, spins in the spin network theory
\cite{fuzz, bfks, qg}.} and that the size of each of these units
is $\stackrel {>} {_\sim} l_f \;$. Since the evolutions are
assumed to be unitary, it then follows that the information
about the star, its collapse, formation of MCO, and its further
evolution may not be lost but must all be encoded among these
units.

The region of singularity resolution may be visualised as a
cloud of fundamental units and, for the sake of convenience
here, will be referred to as `singularity cloud'. As follows
from the above assumptions, its size must depend on mass $M$ of
the collapsing star and must be parametrically much larger than
$l_f \;$. For the purposes of illustration, we present several
plausibile scenario for the number of fundamental units and the
interactions between them, thereby also estimating the size of
the singularity cloud. We will then study its implications in
general.

\vspace{4ex}

\centerline{ \bf A. Size of the singularity cloud: a few
scenario}

\vspace{4ex}

Let $N$ be the number of fundamental units in the singularity
cloud formed when a star of mass $M$ collapses and the
singularities are resolved. This singularity cloud may be taken
to have mass $\sim M$ and an entropy $S_{cld} \;$. Note that, in
general relativity theory, an MCO is believed to be a black hole
with an entropy given by $S_{bh} \sim (l_f \; M)^{\frac{d - 1}{d
- 2}} \; $. On the other hand, the entropy $S_*$ of the star
just before the collapse may be bounded as in \cite{th} by $S_*
\sim (l_f \; M)^{ \frac {d (d - 1)} { (d + 1) (d - 2)}} \;$.
Thus, it is conceivable that the entropy of the resulting
singularity cloud in the MCO is given by $S_{cld} \simeq S_*$
or, equally conceivably, by $S_{cld} \simeq S_{bh} \;$.
\footnote{ There may be other higher entropic possibilities for
$S_{cld}$, as in \cite{page, swz} for example, but they are
not necessary for our purposes here.}

Consider the dependence of the number of fundamental units $N$
on the mass $M \;$. With the values of dimensionless parameters,
if any, all generically set to be of ${\cal O}(1)$, dimensional
analysis implies that the dependence of $N$ on $M$ must be of
the form
\begin{equation}\label{nu}
N \sim (l_f M)^\nu \; \; . 
\end{equation}
Atleast three scenario now suggest themselves, leading to three
different values of the exponent $\nu \;$.

\vspace{2ex}

{\bf (1)} 
The mass of each unit equals $m_f$ and the total mass $M$ equals
the sum of masses of individual units. Then
\begin{equation}\label{nu1}
N = N_1 \sim (l_f M)^{\nu_1} \; \; \; , \; \; \; \; 
\nu_1 = 1 \; \; .
\end{equation}
In this scenario, each unit must have further internal structure
in order to account for the entropy. The number of such internal
degrees of freedom $n_{int}$ for each unit must be of the order
of $n_{int} \sim \frac{S_{cld}} {N_1} \;$. Thus, if $S_{cld} =
S_*$ then $n_{int} \sim (l_f \; M)^{\frac{2}{ (d + 1) (d - 2)}}
\;$ and if $S_{cld} = S_{bh}$ then $n_{int} \sim (l_f \;
M)^{\frac{1}{ d - 2}} \;$.

\vspace{2ex}

{\bf (2)} 
The $N$ units account for the entropy $S_*$ of the star just
before collapse. Then
\begin{equation}\label{nu2}
N = N_2 \sim (l_f M)^{\nu_2} \; \; \; , \; \; \; \; 
\nu_2 = \frac{d (d - 1)}{ (d + 1) (d - 2)} \; \; .
\end{equation}
In this scenario, the average mass $m_{av}$ of each unit is less
than $m_f \;$ and must be of the order of $m_{av} \sim (l_f \;
M)^{- \frac {2} { (d + 1) (d - 2)}} \; \; m_f \;$.

\vspace{2ex}

{\bf (3)} 
The $N$ units account for the entropy $S_{bh}$ of the black hole
that would have formed in general relativity theory. Then

\begin{equation}\label{nu3}
N = N_3 \sim (l_f M)^{\nu_3} \; \; \; , \; \; \; \; 
\nu_3 = \frac{d - 1}{d - 2} \; \; .
\end{equation}
In this scenario, the average mass $m_{av}$ of each unit is less
than $m_f \;$ and must be of the order of $m_{av} \sim (l_f \;
M)^{- \frac {1} { d - 2}} \; \; m_f \;$.

\vspace{2ex} 

Consider now the size of the singularity cloud. This size,
denoted as $L_{cld} \;$, depends on the interactions between the
N fundamental units. The details of these interactions which
arise due to quantum gravity effects are not known. But three
generic physically reasonable possibilities may be considered:
The interactions may be such that the N units may {\bf (a)}
clump together as densely as possible; or {\bf (b)} move
independently of each other, effectively executing N units of
random walks; or {\bf (c)} avoid each other, effectively
executing N units of self avoiding random walks \cite{random}.
These interactions may be thought of as short range and
attractive, neutral, or repulsive.  Given that the size of each
unit is $\stackrel {>} {_\sim} l_f \;$, it follows that the size
of the singularity cloud consisting of $N$ units is given by
\begin{equation}\label{sigma}
L_{cld} \; \stackrel {>} {_\sim} \; N^\sigma \; l_f 
\end{equation}
where the exponent $\sigma$ for the above possible interactions
are given
by \cite{pdg}
\begin{equation}\label{sigmaabc}
\sigma_a \; = \; \frac{1}{d} \; \; , \; \; \;
\sigma_b \; = \; \frac{1}{2} \; \; , \; \; \;
\sigma_c \; = \; max \; \left\{ \frac{3}{d + 2}, \;
\frac{1}{2} \right\} 
\end{equation}
with $d$ being the number of spatial dimensions, taken to be
$\ge 3 \;$ in this paper. Thus, $\sigma_c = \frac{3}{5}$ for $d
= 3$ and $\sigma_c = \sigma_b = \frac{1}{2}$ for $d \ge 4 \;$.
Using equations (\ref{nu}) and (\ref{sigma}), the size of the
singularity cloud $L_{cld}$ is then given in terms of its mass
$M$ by
\begin{equation}\label{alpha}
L_{cld} \; \stackrel {>} {_\sim} \; (l_f M)^\alpha \; l_f 
\; \; \; , \; \; \; \; \alpha = \nu \sigma \; \; . 
\end{equation}
For the various scenario for $N$ dependence on $M$ and $L_{cld}$
dependence on $N$ considerd here, the values of the exponents
$\alpha_{i x} = \nu_i \; \sigma_x \;$ follow easily where $i =
(1, 2, 3)$ and $x = (a, b, c) \;$ label the scenario. These
values are listed in Table 1. Note that $\sigma_c = \sigma_b$
and, hence, $\alpha_{i c} = \alpha_{i b}$ for $d \ge 4 \;$.
Therefore, we have listed $\alpha_{i c}$ for $d = 3$ only.

\vspace{4ex}

\begin{center}

\begin{tabular}{||c||c|c|c||} 
\hline \hline 
Values & & & \\ 
of & a & b & c \\ %\hline \hline 
$\alpha_{i x} = \nu_i \; \sigma_x$ & & & $(d = 3)$ \\ 
\hline \hline
& & & \\ 
1 & $\frac{1}{d}$ 
& $\frac{1}{2}$ 
& $\frac{3}{5}$ 
\\ & & & \\ \hline
& & & \\ 
2 & $\frac{d - 1}{(d + 1) (d - 2)}$ 
& $\frac{d (d - 1)}{ 2 (d + 1) (d - 2)}$
& $\frac{9}{10}$ 
\\ & & & \\ \hline 
& & & \\ 
3 & $\frac{d - 1}{d (d - 2)}$ 
& $\frac{d - 1}{2 (d - 2)}$ 
& $\frac{6}{5}$ \\ & & & \\ \hline \hline
\end{tabular} 

\vspace{2ex}

\noindent {\bf Table 1:} {\em The values of the exponents
$\alpha_{i x} = \nu_i \; \sigma_x \;$ where $i = (1, 2, 3)$ and
$x = (a, b, c) \;$. $\; \sigma_c = \sigma_b$ and, hence,
$\alpha_{i c} = \alpha_{i b} \;$ for $d \ge 4 \;$. Therefore, in
the third column, $\alpha_{i c}$ are listed for $d = 3$ only.}

\end{center}

\newpage 

\vspace{4ex}

\centerline{ \bf III. Implications of $\alpha < \alpha_h$ and
$\alpha > \alpha_h$} 

\vspace{4ex}

Note that, in general relativity theory, the singularity is
taken to be of Planckian size, and the constituents of the
collapsing massive star are taken to have disappeared into this
singularity. One may then say that the size of the singularity
cloud is ${\cal O} (l_f)$ and, hence, $\alpha = 0$ in this
theory. In a quantum theory of gravity, the singularities are
assumed to be resolved by transforming the constituents of the
collapsing massive star into fundamental units. Therefore, the
size of the singularity cloud must depend on its mass and must
be parametrically much larger than $l_f \;$. Hence, the value of
the exponent $\alpha$ in equation (\ref{alpha}) must be non zero
and positive.

Consider now the implications of the values of $\alpha$ being
non zero and positive. In general relativity theory, the horizon
size $r_h$ of a black hole of mass $M$ is given by
\begin{equation}\label{alphah}
r_h \sim (l_f \; M)^{\alpha_h} \; l_f 
\; \; \; , \; \; \; \; \alpha_h = \frac{1}{d - 2} \; \; . 
\end{equation}
Comparing equations (\ref{alpha}) and (\ref{alphah}), it follows
that if $\alpha < \alpha_h$ then the size of the singularity
cloud is smaller than the horizon size of the black hole. This
case can, therefore, be taken to mean that the MCO formed in the
collapse of a massive star has no singularity but has a horizon.
Consequently, it evolves further by emitting Hawking radiation.
Also, because of the horizon, nothing from the singularity cloud
inside the MCO can escape to the outside. The information about
the star, its collapse, formation of MCO, and its further
evolution is encoded among the fundamental units in the
singularity cloud; and, this information is not accessible to an
outside observer.

The horizon shrinks due to the emission of Hawking radiation
and, eventually, becomes smaller than the singularity cloud in
size. This means that the MCO has no horizon now and consists
entirely of the singularity cloud. Consequently, it then evolves
further as a standard quantum system with large number of
interacting degrees of freedom. Objects from the cloud can now
escape to the outside and, hence, the information encoded in the
singularity cloud becomes accessible to an outside observer.
See \cite{g, h1, visser, rv} for a similar scenario .

If $\alpha > \alpha_h$ then it follows that the size of the
singularity cloud is larger than the horizon size of the black
hole. Taken literally, this gives, for example, 
\begin{equation}\label{7.6}
L_{cld} \sim (l_f M)^{\alpha - \alpha_h} \; r_h \; \sim \;
10^{7.6} \; (3 \; km) 
\end{equation}
for a singularity cloud of one solar mass in four dimensional
spacetime if $\alpha = \frac{6}{5} \;$. The size of the MCO is
also likely to be of this order. But this is not the case for a
black hole candidate of one solar mass in our four dimensional
universe. (Its expected size is less than $10 \; km \;$.)

In the collapse of a star in general relativity theory where a
black hole forms, its horizon size is believed to grow from zero
and finally reach $r_h$ at the end of collapse. With the above
assumptions about the singularity resolution and about the value
of $\alpha$ being $> \alpha_h \;$, what may likely happen in
such a collapse in a quantum theory of gravity is that the
singularity cloud will also grow from zero size, but always
remaining larger than the corresponding horizon only by a factor
of order unity, and finally reach a size larger than the
predicted horizon size by a similar factor of order unity.
\footnote{ This may be the case if, for example, $\alpha =
\alpha_h + \epsilon$ where $\epsilon$ is a small positive
number. In the limit $\epsilon \to 0_+ \;$, the size of the
singularity cloud may be given by $L_{cld} \; \sim \; (l_f
M)^{\alpha_h} \; ln (l_f M) \; l_f \; \sim \; r_h \; ln (l_f M)
\;$.} Clearly, a detailed theory is needed to predict
dynamically how a collapse proceeds, how the fundamental units
are produced and generate the singularity cloud, and finally to
predict the size of the resulting MCO.

Throughout in this paper, therefore, we take the $\alpha >
\alpha_h$ case to mean only that $L_{cld} > r_h$ and that the
MCO formed in the collapse of a massive star has no singularity
and also has no horizon, and consists entirely of the
singularity cloud. Consequently, it evolves further as a
standard quantum system with large number of interacting degrees
of freedom. Objects from the cloud can escape to the outside
and, hence, the information about the star, its collapse,
formation of MCO, and its further evolution encoded among the
fundamental units in the singularity cloud remains accessible to
an outside observer at all times.

If $\alpha = \alpha_h$ then $L_{cld} \sim r_h \;$. The exact
coefficient depends on the details of the quantum theory of
gravity. Hence we will not consider this case and, instead, take
that the $L_{cld} < r_h \;$ case is covered by $\alpha <
\alpha_h$, and that the $L_{cld} > r_h \;$ case is covered by
$\alpha > \alpha_h \;$. See also footnote 3.

Returning to the values of $\alpha$ listed in Table 1 for
various scenario, note that $\nu_1 < \nu_2 < \nu_3$ and, hence,
$\alpha_{1 x} < \alpha_{2 x} < \alpha_{3 x}$ for $x = (a, b, c)
\;$. Also, $\alpha_h = \frac {1} {d - 2} \;$. Hence if $x = a
\;$, namely if the interactions are such that the N fundamental
units clump together as densely as possible, then
\begin{equation}\label{ge3} 
\alpha_{1 a} < \alpha_{2 a} < \alpha_{3 a} < \alpha_h
\end{equation}
for $d \ge 3 \;$. If $x = b \;$, namely if the interactions are
such that the N fundamental units move independently of each
other and effectively execute N units of random walks, then 
\begin{equation}\label{e3b} 
\alpha_{1 b} < \alpha_{2 b} < \alpha_{3 b} = \alpha_h
\end{equation}
for $d = 3 \;$ and 
\begin{equation}\label{ge4} 
\alpha_h \le \alpha_{1 b} < \alpha_{2 b} < \alpha_{3 b}
\end{equation}
for $d \ge 4 \;$. If $x = c \;$, namely if the interactions are
such that the N fundamental units avoid each other and
effectively execute N units of self avoiding random walks, then
\begin{equation}\label{e3c} 
\alpha_{1 c} < \alpha_{2 c} < \alpha_h < \alpha_{3 c} 
\end{equation}
for $d = 3 \;$. For $d \ge 4 \;$, we have $\alpha_{i c} =
\alpha_{i b}$ and equation (\ref{ge4}) applies. 

Note, in particular, that $\alpha_{i a} < \alpha_h \;$ for all
$i \;$ and that $\alpha_{3 c} > \alpha_h \;$. Thus, if we assume
that the $N$ fundamental units clump together as densely as
possible ($x = a$) then we have $\alpha < \alpha_h \;$. In these
scenario, therefore, the collapse of a massive star is likely to
result in the formation of an MCO which has no singularity but
has a horizon. If we assume that the $N$ fundamental units
account for the entropy of the black hole that would have formed
in general relativity theory ($i = 3$) and that these $N$ units
execute self avoiding random walks ($x = c$) then we have
$\alpha > \alpha_h \;$. In these scenario, therefore, the
collapse of a massive star is likely to result in the formation
of an MCO which has no singularity and also has no horizon, see
the comments below equation (\ref{7.6}).

\vspace{4ex}

\centerline{ \bf IV. Consequence of $0 < \alpha < \alpha_h \;$:
Loss of predictability }

\vspace{4ex}

Among the values of $\alpha$ listed in Table 1, $\alpha_{1 a}$
is the smallest. This corresponds to the most conservative
assumptions, namely that the mass of each fundamental unit is
$m_f$, which leads to the smallest number $N \sim l_f M$, and
that these $N$ units clump together as densely as possible,
which leads to the smallest size $L_{cld} \sim (l_f M)^{\frac
{1} {d}} \; l_f \;$. This size has also been obtained by others.
See \cite{ morgan, g, rv} for example, and the references
therein.

We now assume only that $\alpha$ is strictly positive and study
what happens if $\alpha < \alpha_h \;$, see \cite{k12} for an
earlier report. The collapse of a massive star in this case is
likely to result in the formation of an MCO which has no
singularity but has a horizon. Let $M_{init}$ be its mass at the
time of formation. Then the size of the singularity cloud $\sim
(l_f M_{init} )^\alpha \; l_f \;$ and, since $\alpha > 0 \;$, it
is parametrically much larger than $l_f \;$. The singularity
cloud consists of $N_{init}$ number of fundamental units, among
which the information about the star and its collapse is
encoded.

Consider the evolution of this MCO after formation. It may
accrete more mass which will increase its mass and its horizon
size. The accreted mass will also increase the size of the
singularity cloud and the number of fundamental units in it.

Since the MCO has a horizon, it evolves further by emitting
Hawking radiation: a pair of photons is created at the horizon;
one of them, the out-photon, goes outside the horizon and the
other one, the in-photon, falls inside; the in-photon has
negative energy and, hence, it reduces the mass of the MCO and
shrinks its horizon size.

In the present case, where the singularities are assumed to be
resolved, the in-photons will fall into the singularity cloud.
They carry negative energy, so the nett mass of the singularity
cloud should decrease. Interacting with the fundamental units in
the singularity cloud, the in-photons are likely to generate
some `decay products' and form a kind of `plasma' in
equilibrium.\footnote{ This is analogous to negative charges
falling into a collection of positive charges and producing
decay products, all confined in a region. The negative charges
annihilate some positive charges and produce photons which are
also confined in the region. These photons, in turn, produce
pairs of negative and positive charges. In equilibrium then,
there will be a plasma of negative and positive charges together
with photons, the decay products.}

In particular, the in-photons must not decrease the number of
fundamental units by, for example, disappearing completely
together with some fundamental units leaving nothing behind. On
the contrary, the in-photons must increase the number of
fundamental units and, thereby, increase the size of the
singularity cloud. This is because the fundamental units encode
information about the star, its collapse, formation of MCO, and
its further evolution. If the in-photons can decrease the number
$N$ of fundamental units by disappearing completely together
with some of them then, in principle, $N$ can decrease by
arbitrary amounts if, for example, Hawking radiation lasts for
arbitrarily long time with accretions suitably compensating the
resulting mass loss. In such cases, there will not be sufficient
number of fundamental units available in the singularity cloud
to encode all the information about the star, its collapse,
formation of MCO, and its further evolution through Hawking
radiation and more accretions. This will then contradict the
assumption of unitary evolution since information can not be
lost in such an evolution.

It therefore follows that the in-photons of the Hawking
radiation must increase the number of fundamental units in the
singularity cloud and, thereby, increase its size. \footnote {A
similar result on the increase in size is also arrived at in
\cite{h1, visser, rv} where the surface of the region of
singularity resolution is modelled as an inner horizon.}

Now, imagine observing today an MCO of mass $M_{today} \;$.
Thinking of it as a black hole or, more precisely, as a black
hole candidate, its horizon size is given by
\begin{equation}\label{rhtoday} 
r_{h \; (today)} \sim (l_f \; M_{today})^{\frac{1}{d - 2}} \;
l_f \; \; \; .  
\end{equation}
With the singularities assumed to be resolved, there is a
singularity cloud inside this black hole candidate. Its size
$L_{cld \; (today)}$, however, depends on {\bf (i)} the initial
mass $M_{init}$ of the collapsing star, {\bf (ii)} accreted mass
$M_{accrn}$ till today, and {\bf (iii)} the amount of Hawking
radiation emitted till today. From the above discussions, it
follows that the size of the singularity cloud obeys the
inequality
\begin{equation}\label{lcldtoday} 
L_{cld \; (today)} \; > \; 
(l_f \; (M_{init} + M_{accrn}))^\alpha \; l_f
\; \; \; , \; \; \; \; \; \; \alpha > 0 \; \; , 
\end{equation}
and need not have any relation to the observed mass $M_{today}$
or to the horizon size $r_{h \; (today)} \;$. This means that
the future evolution of the black hole observed today is
uncertain. We cannot predict when the singularity cloud will
become larger than the horizon in size, after which objects from
it can escape to the outside and, hence, information about the
original star, its collapse, and its further evolution encoded
among the fundamental units in the singularity cloud will become
accessible to an outside observer. Thus there is a loss of
predictability. See Appendix A for a brief explanation regarding
the meaning of the word predictability as used here.

This loss of predictabilty is not there in general relativity
theory. This is because, in this theory, the size of the
singularity is always taken to be ${\cal O} (l_f) \;$,
equivalently $\alpha = 0 \;$, and it has no dependence on
$M_{init}$, or on $M_{accrn}$, or on the amount of Hawking
radiation emitted. It therefore seems that the resolution of
singularity has led us to a worse predicament !

This loss of predictabilty is, again, not there if $\alpha >
\alpha_h \;$ because it then follows that the size of the
singularity cloud is larger than the horizon size of the black
hole. This means that the MCO has no horizon and that all the
information about the star, its collapse, formation of MCO, and
its further evolution remains accessible to an outside observer
at all times.

Thus, if $0 < \alpha < \alpha_h$ then there is a loss of
predictability described above. The resolution of singularity
then seems to lead to a worse predicament than in general
relativity theory. The singularity resolution is expected in a
quantum theory of gravity, but not the loss of predictability.
Therefore, if we further assume that a quantum theory of gravity
must not lead to the loss of predictability then it logically
follows that the physics of singularity resolution must be such
that $\alpha > \alpha_h \;$. As explained earlier below equation
(\ref{7.6}), this inequality is to be taken to mean only that
the size of the singularity cloud is larger than the horizon
size of the black hole.

%\newpage 

\vspace{4ex}

\centerline{ \bf V. Summary and Conclusion}   

\vspace{4ex}

In summary, we considered the collapse of a massive star forming
an MCO in a quantum theory of gravity. We followed a
qualitative, physically motivated approach since a suitable
quantitative one is not available. We assumed the following.

\begin{itemize}

\item

A quantum theory of gravity exists where the singularities are
resolved and the evolutions are unitary.

\item

The mass $M$ of the collapsing star is $\gg m_f \;$. 
Dimensionless parameters, if any, all take generic values
of ${\cal O}(1)$ independently of $M \;$.

\item

The singularities are resolved by transforming the star's
constituents into fundamental units, each of whose size is
$\stackrel {>} {_\sim} l_f \;$.

\item

A quantum theory of gravity must not lead to the loss of
predictability.

\end{itemize}

The above assumptions are physically well motivated. The first
three assumptions imply the following.

\begin{itemize}

\item 

The information about the star, its collapse, formation of MCO,
and its further evolution may not be lost but must all be
encoded among the fundamental units.

\item 

The size of the singularity cloud must depend on mass $M$ of the
collapsing star and must be parametrically much larger than $l_f
\;$. Hence, $\alpha$ in equation (\ref{alpha}) must be non zero
and positive.

\end{itemize}

We then argued that the fourth assumption implies the following.

\begin{itemize}

\item

$\alpha$ must be $> \alpha_h \;$ in the sense that the size of
the singularity cloud is larger than the horizon size of the
black hole.

\end{itemize}

Therefore, we conclude that in a quantum theory of gravity
satisfying the assumptions listed above, the nature of an MCO
formed in the collapse of a massive star must be as follows: It
has no singularity and also has no horizon. It evolves further
as a standard quantum system with large number of interacting
degrees of freedom. And, the information about the star, its
collapse, et cetera remains accessible to an outside observer at
all times.

The arguments presented in this paper and, hence, the resulting
conclusions are very general. They are applicable to any quantum
theory of gravity satisfying the present assumptions. The flip
side of this generality is that the present arguments do not
give any hint of the mechanisms responsible, for example, for
the singularity resolutions, or for the production of
fundamental units, or for the size of the singularity cloud. The
nature and the details of such mechanisms are strongly dependent
on the theory considered. Clearly, a complete knowledge of the
underlying theory is needed to understand them in detail. These
details, in turn, are crucial for applications to the actual
collapse of stars, and to predict the size and the properties of
the resulting MCOs.

We close by mentioning two studies which are fruitful and are
likley to provide more insights. In general relativity, the MCOs
are believed to be black holes, the simplest ones being
described by the Schwarzschild solutions. In Brans -- Dicke
theory, for example, there are more general Janis -- Newman --
Winicour -- Wyman (JNWW) solutions \cite{jnww}. They have
singularities but no horizon at the Schwarzschild radius
$r_{sch} \;$. Because of these singularities, they are not
expected to describe an MCO. However, similar singular solutions
generically arise in many contexts including, for example, in
higher dimensional contexts relevant for string/M theory
\cite{k13a}. In a quantum theory of gravity where the
singularities are assumed to be resolved, one may expect such
JNWW-type solutions to describe MCOs. With singularities
resolved, these objects are likely to have no horizon and their
sizes are likely to be $\stackrel {>} {_\sim} \; r_{sch}
\;$. Indeed, such singularity-free, horizonless solutions, with
Kaluza -- Klein monopoles decorating a surface of ${\cal
O}(r_{sch})$ radius appear in the fuzz ball descriptions in
string theory, see \cite{kkmonopole, fuzz} and references
therein. It is important to study such JNWW-type solutions with
their singularities resolved by quantum gravity effects.

Another aspect that can be fruitfully studied with the presently
available techniques is the stability properties of the MCOs
which have no singularities and also have no horizon. The
arguments of this paper would imply that adding more mass to
such an MCO should result in the increase of its size but no
collapse should be possible; hence, there should be no
instabilities towards a collapse irrespective of how massive an
MCO is. See \cite{k13b} for a preliminary study of this aspect
in the context of string/M theory. It is not clear what
properties of the constituents of an MCO such as their equations
of state, their description as multi perfect fluids,
interactions between them, et cetera will ensure the stabilty
against collapse.  Nevertheless, such a study seems possible
even in the absence of a detailed quantum theory of gravity
applicable to a collapsing star, and seems likely to provide
useful insights.
 
\vspace{3ex}

{\bf Acknowledgement:} 
We thank B. Sathiapalan for discussions. 

\newpage

\vspace{4ex}

\centerline{\bf Appendix A. Regarding the meaning of loss of
predictability}

\vspace{4ex}

In this paper, we use the word predictability to mean only the
ability to predict the future evolution of a system by making
measurements now, and without requiring to know the past history
of the system and its constituents. This actually seems to be
more a statetment about the nature of the theories one
constructs and tests in physics. A detailed discussion of these
aspects is beyond the scope of the present paper. Hence,
instead, we give below two examples of systems which are
predictable in the sense used here.

A box of gas is a predictable system. We can measure the values
of a suitable set of observables and predict the future
evolution of the gas in the box to a specified accuracy. The
past histories of gas molecules are not needed.

Schwarzschild black hole in semiclassical gravity is
predictable. If its mass is known then Hawking's calculations
tell us that it will emit radiation and evaporate to nothing in
a time $\propto \; (mass)^3 \;$ in four dimensional spacetime.
The past history of the black hole from the time of its
formation is not needed. Knowing its present mass alone is
sufficient.

In the $0 < \alpha < \alpha_h$ case considered here, there is a
loss of predictability in this sense. The past history of the
MCO from the time of its formation is needed to predict its
future evolution. Knowing its present mass alone is not
sufficient. Hence, this system is not predictable in the sense
used here.

It is perhaps debatable whether such a loss of predictability is
undesireable. In this paper, we assume that such a loss is
undesireable since the physical theories we are familiar with
are of the predictable type.


\begin{thebibliography}{999}
\bibitem{random}
One may imagine the space to have a lattice structure in the
region of singularity resolution. The dynamics of singularity
resolution by transforming the constituents of the star into
fundamental units may then be such that the resulting $N$ units
are, in an average sense, centered around $N$ spatial lattice
sites that a random walker might have visited.
\end{thebibliography}
\end{document}